\documentclass[11pt]{article}
\usepackage{amsmath}
\usepackage{amssymb}
\usepackage{amsfonts}
\usepackage{eucal}
\usepackage{epsfig}
\newcommand{\be}{\begin{equation}}
\newcommand{\bea}{\begin{eqnarray}}
\newcommand{\eea}{\end{eqnarray}}
\newcommand{\ba}{\begin{array}}
\newcommand{\ea}{\end{array}}
\newcommand{\ee}{\end{equation}}

\expandafter\ifx\csname mathbbm\endcsname\relax

\else

\fi

\renewcommand{\baselinestretch}{1}

\begin{document}

\title{Collective coordinate variable for soliton-potential system in sine-Gordon model}  \author{Kurosh Javidan
\\javidan@um.ac.ir  \\Department of physics, Ferdowsi university of Mashhad\\ 91775-1436 Mashhad Iran }
\setlength{\parindent}{0mm} 
\renewcommand{\baselinestretch}{1} 
\newcommand{\ph}{\vec{\phi}} \newcommand{\pha}{\phi_{a}}
\newcommand{\dmu}{\partial_{\mu}} \newcommand{\umu}{\partial^{\mu}}
\newcommand{\dnu}{\partial_{\nu}} \newcommand{\unu}{\partial^{\nu}}
\newcommand{\di}{\partial_{i}} \newcommand{\ui}{\partial^{i}}
\renewcommand{\dj}{\partial_{j}} \newcommand{\uj}{\partial^{j}} \hoffset =
-2cm \textwidth = 170mm
\maketitle
\abstract

A Collective coordinate variable for adding a space dependent potential to the sine-Gordon model is presented. Interaction of solitons with a delta function potential barrier and also delta function potential well is investigated. Most of the characters of interaction are derived analytically. We will find that the behaviour of a solitonic solution is like a point particle which is moved under the influence of a complicated effective potential. The effective potential is a function of the field initial conditions and also parameters of added potential. 

\section{Introduction}

Solitons are important objects in most of the branches of sciences. They are stable against dispersive effects, and propagate very similar to classical point-like particles.

Topological solitons play important role in non-perturbative aspects of quantum field theories. They are widely used as models of particles which are generated as nontrivial solutions of nonlinear field theories. Skyrmions are solitons which are used as a model of hadrons. Solitons also widely appear in models of condensed matter, such as Josephson junction, fluxon, spin chain dynamics and so on. Describing active regions of DNA is another example of soliton applications in other branches of sciences \cite {rr1, rr2, rr3}.

Sine-Gordon equation is a very famous model with well known topological soliton solutions. This model widely appears in nonlinear problems. Some of the above mentioned situations are examples of sine-Gordon applications. 

Recently, there has been an increasing interest in the scattering of solitons from defects or impurities, which generally come from medium properties. The motivations come from both the theoretical and the application part of physics. The effects of medium disorders and impurities can be added to the equation of motion as perturbative terms \cite {r1,r2}. These effects also can be taken into account by making some parameters of the equation of motion to be function of space or time \cite {r3,r4}. There still exists another interesting method which is mainly suitable for working with topological solitons \cite {r5, r6}. In this method, one can add such effects to the Lagrangian of the system by introducing a suitable nontrivial metric for the back ground space-time, without missing the topological boundary conditions . 

Interaction of solitons with defects mainly investigates using numerical analysis and numerical simulations. Some analytical models have been presented which are constructed with using suitable collective coordinate systems. Analytical models for the sine-Gordon and $\phi^{4}$ field theories has been presented base on the method of adding the defects through the background space-time metric \cite {r7, r8}. They predict most of the "soliton-potential" behaviours with a very good precision. This method can be used for objects that their equation of motion results from a Lorentz invariant action, such as sine-Gordon model, $\phi^{4}$  theory, $CP^{N}$  model, Skyrme model, Faddeev-Hopf equation, chiral quark-soliton model, Gross-Neveu model, nonlinear Klein-Gordon models and so on.

In this paper a simple collective coordinate system is presented. This system is constructed base on the method of adding the defects by making some parameters of Lagrangian to be function of space. This method can be used for other nonlinear field theories which their equations of motion are not Lorentz invariant as well as Lorentz invariant field theories.

Therefore a new collective coordinate system for solitons of the sine-Gordon field theory will be presented in section 2. Results of this model are discussed and also they will compared with results of Ref. \cite {r7} in section 3. Some conclusion and remarks will be presented in final section 4.


\section{Collective coordinate variable}

Sine-Gordon model in (1+1) dimensions $\mu$=0,1, is defined by

\begin{equation}\label{lag}
{\cal L}=\frac{1}{2}\partial_ {\mu}\phi\partial^{\mu}\phi-\lambda(x)\left(1-\cos\phi\right)   
\end{equation}
where $\lambda(x) = \lambda_{0} + V(x)$. V(x) is a potential parameter and carries the effects of the external potential. Potential V(x) is a localized function which is nonzero only in a certain region of space.

The equation of motion for the Lagrangian (\ref{lag}) is :
\begin{equation}\label{eq}
\partial_ {\mu}\partial^{\mu}\phi+\lambda(x)\sin(\phi)=0   
\end{equation}
This equation has not analytical solution with a general function for V(x). If we take V(x)=0, we have usual sine-Gordon equation with the following one soliton solution
\begin{equation}\label{sol}
\phi (x,X(t))= 4tan^{-1}\left(\exp \left(\sqrt{\lambda_{0}}\frac{x-X(t)}{\sqrt{1-\dot{X}^2}}   \right)\right)
\end{equation}
where $X(t)=x_{0}-\dot{X}t$. $x_{0}$ and $\dot{X}$ are soliton initial position and its velocity. In the following calculations $\lambda_{0}$ has been set to $\lambda_{0}=1$ . 

Center of the soliton can be considered as a particle if we look at this as a collective coordinate variable. The collective coordinate could be related to the potential by using suitable function V(x) in equation (\ref{eq}). Thus the model is able to give us analytic description for the evolution of the soliton center during the soliton-potential interaction. By inserting the solution (\ref{sol}) in the Lagrangian (\ref{lag}) with adiabatic approximation \cite {r1,r2} we have
\begin {equation} \label {le1}
{\cal L}=2\left(\dot{X}^{2}-1\right)sech^{2}\left(x-X(t)\right)-2\lambda(x)sech^{2}\left(x-X(t)\right) 
\end {equation}
X(t) remains as a collective coordinate if we integrate (\ref{le1}) over the variable x.
\begin {equation} \label {L}
L=\int{{\cal L}dx}=4\dot{X}^{2}-8-2\int{V(x)sech^{2}\left(x-X(t)\right)dx}
\end {equation}
It is a general equation for any kinds of external potential V(x). If we take the potential $V(x)=\epsilon \delta(x)$ (\ref{L}) becomes
\begin {equation} \label {LL}
L=4\dot{X}^{2}-8-2\epsilon sech^{2}\left(x-X\right)
\end {equation}
The equation of motion for the variable X(t) is derived from (\ref{LL})
\begin {equation} \label {EqX}
8\ddot{X}-4\epsilon sech^{2}(X)\tanh(X)=0
\end {equation}
 
The above equation shows that the peak of the soliton moves under the influence of a complicated force which is a function of soliton position, soliton velocity and characters of external potential $V(x)$. If $\epsilon>0$ we have a barrier and $\epsilon<0$ creates a potential well. Equation (\ref{EqX}) has an exact solution as follows
\begin {equation} \label {Xdot}
\dot{X}^2=\frac{1}{2}\epsilon sech^{2}(X_{0})+\dot{X}_{0}^{2}-\frac{1}{2}\epsilon sech^{2}(X)
\end {equation}
where $X_{0}$ and $\dot{X}_{0}$ are soliton initial position and its initial velocity respectively. 

Hamiltonian density is obtainable from Lagrangian (\ref{le1}) as follows
\begin {equation} \label {ham}
{\cal H}=2\left(\dot{X}^{2}+1\right)sech^{2}\left(x-X(t)\right)+2\lambda(x)sech^{2}\left(x-X(t)\right) 
\end {equation}
Collective energy density can be calculated with integration of (\ref{le1}) over the variable x. Energy of the soliton in the presence of the potential $V(x)=\epsilon \delta(x)$ becomes
\begin {equation} \label {Energy}
E=4\dot{X}^2+8+2 \epsilon sech^{2}X
\end {equation}
It is the energy of a particle with the mass of m=8 and velocity $\dot{X}$ which is moved under the influence of external effective potential. Figure 1 presents energy of a static soliton as a function of its position under the influence of the potential $V(x)=\epsilon \delta(x)$ with $\epsilon =0.5$. Because of the extended nature of the soliton, the effective potential is not an exact delta function.  
\begin{figure}[htbp]
  \begin{center}
    \leavevmode
 \epsfxsize=11cm   \epsfbox{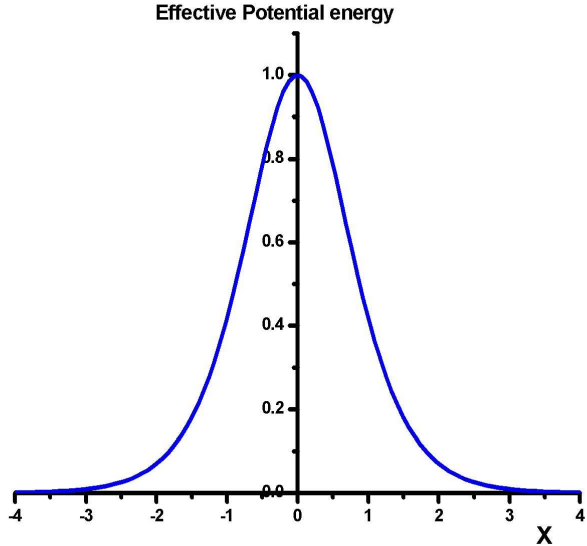}
  \end{center}
  \caption{Energy of a static soliton as a function of its distance from the potential $V(x)=\epsilon \delta(x)$ with $\epsilon=0.5$. }
  \label{fig:fig1}
\end{figure}

By substituting $\dot{X}$ from (\ref{Xdot}) into (\ref{Energy}) one can show that the energy is a function of soliton initial conditions $X_{0}$ and $\dot{X}_{0}$ only. Therefore the energy of the system is conserved. 

Topological charge is
\begin {equation} \label {Q}
Q=\int^{-\infty}_{+\infty}{\frac{\partial\phi}{\partial x}dx}
\end {equation}
where $\frac{\partial\phi}{\partial x}$ is topological charge density. For the solution (\ref{sol}) we have 
\begin {equation} \label {q}
\frac{\partial\phi}{\partial x}=2sech(x-X)
\end {equation}
Collective topological charge density q(X) can be calculated by Integrating (\ref{q}) over the variable x. A simple calculation shows that it is constant and independent of the soliton position 'X'. 

Some features of soliton-potential dynamics can be investigated using equations (\ref{Xdot}) and (\ref{Energy}) analytically which are discussed in the next section.

\section{Soliton-potential dynamics }
A soliton-barrier system is modeled with $\epsilon>0$ in Eq. (\ref{Xdot}) or Eq. (\ref{Energy}). Consider a soliton which is placed far away from the potential (namely at infinity), with initial velocity $\dot{X}_{0}$. It moves toward the barrier and interacts with it. There exist two different kinds of trajectories for the soliton after the interaction with the barrier, depend on its initial velocity, which separate by a critical velocity $u_{c}$. In low velocities , soliton reflects back and reaches its initial place with final velocity $u_{f}\approx -\dot{X}_{0}$. A soliton with an initial velocity $\dot{X}_{0}>u_{c}$ has enough energy for climbing the barrier, and passing over the potential. At the velocities $\dot{X}_{0}\approx u_{c}$ soliton interacts with the potential slowly and spends more times near the barrier, but this situation is not a bound state. Presented model predicts all of the above features. 

Energy of a soliton in the origin is calculated from (\ref{Energy}) as $E(X=0)=4\dot{X}^{2}+8+2\epsilon$. Minimum value of the energy for a soliton in this situation is $E_{min}(X=0)=8+2\epsilon$. On the other hand, a soliton which comes from the infinity with initial velocity $\dot{X}_{0}$ has the energy of $E(X_{0}=\infty)=4\dot{X}_{0}^{2}+8$ . It is clear that it can pass though the barrier if $E(X_{0}=\infty)\geq E_{min}(X=0)$. Therefore we have
\begin {equation} \label {ucinf}
u_{c}\left({X_{0}=\infty}\right)=\sqrt{\frac{\epsilon}{2}}
\end {equation}
Now consider a soliton which is placed at initial position $X_{0}<0$ (which is not necessarily infinity) with initial velocity $\dot{X}_{0}$. Equation (\ref{Xdot}) shows that the soliton reaches $-\infty(+\infty)$ with final speed $\dot{X}(\infty)=\sqrt{\dot{X}_{0}^2+\frac{1}{2}\epsilon sech^2\left(X_{0}\right)}$ if its initial velocity is less (more) than the critical velocity $u_{c}(X_{0})$. Critical velocity in this situation is not $\sqrt{\frac{\epsilon}{2}}$. This soliton has the critical initial velocity if its initial energy becomes equal to the energy of static soliton at top of the barrier X=0.In this situation the critical velocity is
\begin {equation} \label {ucx0}
u_{c}\left(X_{0}\right)=\sqrt{\frac{\epsilon}{2}\left(1-sech^2(X_{0}) \right)}
\end {equation}

\begin{figure}[htbp]
  \begin{center}
    \leavevmode
 \epsfxsize=11cm   \epsfbox{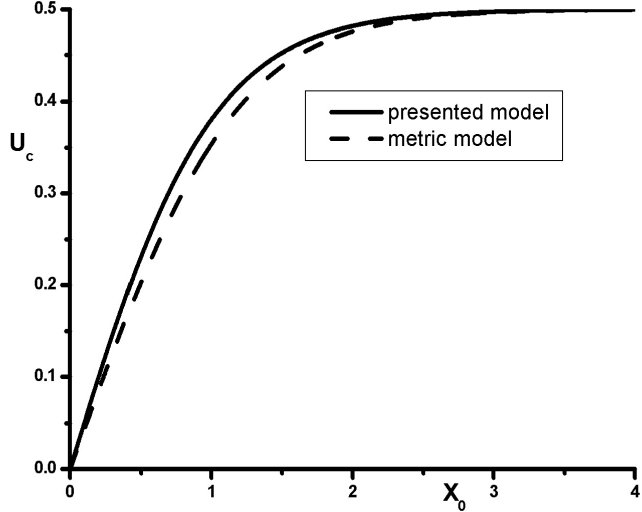}
  \end{center}
  \caption{Critical velocity as a function of initial position predicted in Ref. \cite {r7} (dash line) and our presented model Eq. (\ref{ucx0}) (solid line). }
  \label{fig:fig2}
\end{figure}
All presented models predict same critical velocity for a soliton at the infinity. But critical velocity of a soliton at initial position $X_{0}$ is model dependent. For example following expression has been calculated in \cite {r7}
\begin {equation} \label {ucx02}
u_{c}\left(X_{0}\right)=\sqrt{\frac{\epsilon}{2}\frac{cosh^{2}X_{0}-1}{cosh^{2}X_{0}+\frac{3\epsilon}{4}}}
\end {equation}
Critical velocity of Ref. \cite {r7} (Eq.(\ref{ucx02})) is compared with (\ref{ucx0}) in Figure 2 with $\epsilon=0.5$. Results of both two models are the same at $X_{0}=\infty$ and $X_{0}=0$. But transient behaviour of two models between these limits are different. It is clear that we can not fit one of the models on each other by choosing an effective potential. 

There exists a return point if soliton initial velocity is less than the critical velocity. This point can be found by comparing the soliton energy at its initial position and somewhere that its velocity becomes zero. Therefore we have   
\begin {equation} \label {xstop}
\frac{1}{\cosh^{2}X_{stop}}=\frac{2}{\epsilon}\dot{X}^2_{0}+\frac{1}{\cosh^{2}X_{0}}
\end {equation}
Equation (\ref{xstop}) clearly shows a linear relation between $\frac{1}{\cosh^{2}X_{stop}}$ and $\frac{1}{\cosh^{2}X_{0}}$. There is another linear relation between $\frac{1}{\cosh^{2}X_{stop}}$ and $\frac{1}{\epsilon}$. Direct simulation and also presented analytic model of \cite {r7} confirm these relations too.    

A potential well can be constructed with $\epsilon<0$. Consider a particle moves toward a frictionless potential well from infinity. It falls in the well with an increasing velocity and reaches the bottom of the well with its maximum speed. After that, it will climb the well with decreasing velocity and finally pass through the well. Its final velocity after the interaction is equal to its initial speed. A soliton-well system is constructed with changing $\epsilon$ into $-\epsilon$ in the above equations. But dynamics of a soliton-well system is very different from what we have seen in soliton-barrier interaction. There is not a critical velocity for a soliton-well system, but we can define an escape velocity. A soliton at initial position $X_{0}$ reaches the infinity with a zero final velocity if its initial velocity is
\begin {equation} \label {Xdots}
\dot{X}_{escape}=\sqrt{\frac{\epsilon}{2}sech^{2}\left(X_{0}\right)}  
\end {equation}
Predicted value for this situation in Ref.\cite {r7} is
\begin {equation} \label {Xdots2}
\dot{X}_{escape}=\sqrt{\frac{\epsilon}{2}\frac{1}{\cosh^{2}X_{0}-\frac{3\epsilon}{4}}}  
\end {equation}
Figure 3 presents $\dot{X}_{escape}$ as a function of soliton initial position using (\ref{Xdots}) and (\ref{Xdots2}). Predicted $\dot{X}_{escape}$ at the center of the potential in two models is different. But one can show that we can fit this value by defining an effective potential parameter $\epsilon_{effective}$, however the transient behaviour between the origin and infinity can not fitted. figure 3 also show that predicted escape velocity for infinity in two models is zero. 
\begin{figure}[htbp]
  \begin{center}
    \leavevmode
 \epsfxsize=11cm   \epsfbox{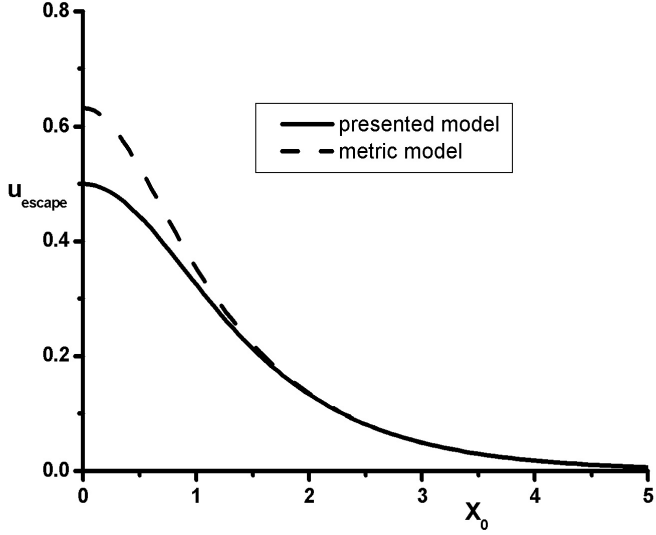}
  \end{center}
  \caption{Escape velocity as a function of initial position predicted in Ref. \cite {r7}, Eq. (\ref{Xdots2}) (dash line) and our calculation Eq. (\ref{Xdots}) (solid line). }
  \label{fig:fig3}
\end{figure}

Consider a potential well with the depth of $\epsilon$ and a soliton at the initial position $X_{0}$ which moves toward the well with initial velocity $\dot{X_{0}}$ smaller than the $\dot{X}_{escape}$. The soliton interacts with the potential well and reaches a maximum distance $X_{max}$ from the center of the potential with a zero velocity and then come back toward the well for another interaction. The soliton oscillates around the well with amplitude $X_{max}$. 
\begin {equation} \label {Xmax}
X_{max}=sech^{-1} \sqrt{sech^{2}X_{0}-\frac{2}{\epsilon}\dot{X}_{0}^{2}}  
\end {equation}
Maximum distance has been calculated with another model in \cite {r7} as follows
\begin {equation} \label {Xmax2}
X_{max}=sech^{-1} \sqrt{sech^{2}X_{0}-\frac{2}{\epsilon}\dot{X}_{0}^{2}\left(1-\frac{3\epsilon}{4}sech^{2}X_{0}\right)}  
\end {equation}
Figure 4 presents $sech^{2}X_{max}$ as a function of $X_{0}$ using (\ref{Xmax}) and (\ref{Xmax2})with $\epsilon=0.5$ and $\dot{X}_{0}=0.5$.They are very similar to each other and it is possible to fit one of them on the other in the origin, with defining an effective potential parameter $\epsilon_{effective}$.
\begin{figure}[htbp]
  \begin{center}
    \leavevmode
 \epsfxsize=11cm   \epsfbox{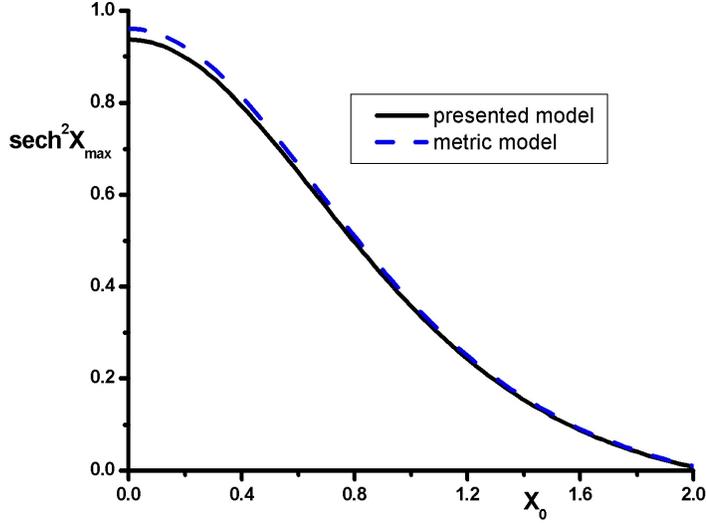}
  \end{center}
  \caption{$sech^{2} X_{max}$ as a function of initial position predicted in Ref. \cite {r7}, Eq. (\ref{Xmax2}) (dash line) and our calculation Eq. (\ref{Xmax}) (solid line). }
  \label{fig:fig4}
\end{figure}

Trajectory of a soliton during the interaction with the potential,X(t) follows from (\ref{Xdot}) as
\begin {equation} \label {t}
t=\int^{X(t)}_{X(t=0)}\frac{dx}{\sqrt{\dot{X}_{0}^{2}+\frac{\epsilon}{2}sech^{2}X_{0}-\frac{\epsilon}{2}sech^{2}X}}
\end {equation}
for soliton-barrier ($\epsilon>0$) and soliton-well ($\epsilon<0$) systems.
The metric model predicts following complicated equation \cite {r7}
\begin {equation} \label {t1}
t=\int^{X(t)}_{X(t=0)}\frac{dx}{\sqrt{\left(\dot{X}_{0}^2 +\frac{2}{3}\right)\frac{\cosh^{2}X \left(\frac{3 \epsilon}{4}+\cosh^{2} X_{0} \right)}{\cosh^{2}X_{0}\left(\frac{3 \epsilon}{4}+ \cosh^{2} X \right)}-\frac{2}{3}}}
\end {equation}
Figure 5 presents soliton trajectory respect to time for a soliton with initial velocity $\dot{X}_{0}=0.1$ and initial position $X=-3$ in a potential well of $\epsilon=-0.3$, simulated with (\ref{t}) and (\ref{t1}). Figure 5 clearly shows very good agreement between presented model and metric model in Ref. \cite {r7} in low velocities. 
\begin{figure}[htbp]
  \begin{center}
    \leavevmode
 \epsfxsize=11cm   \epsfbox{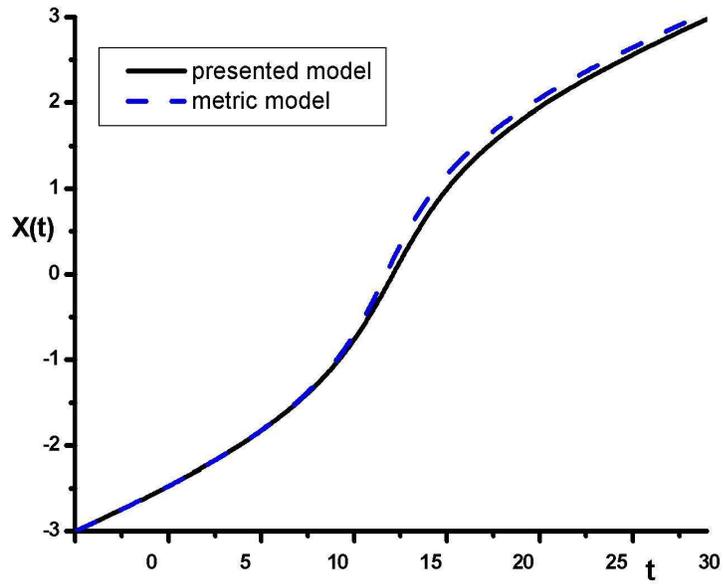}
  \end{center}
  \caption{Soliton trajectory as a function of time for a soliton with initial velocity $\dot{X}_{0}=0.1$ and initial position $X=-3$ during the interaction with potential well $\epsilon=-0.3$ simulated with presented model Eq.(\ref{t})(solid line) and metric model of Ref. \cite {r7} Eq.(\ref{t1}) (dash line). }
  \label{fig:fig5}
\end{figure}

If soliton initial velocity is lower than the escape velocity the soliton oscillates around the well. The period of oscillation can be calculated numerically using equation (\ref{t}). Some simulations have been done using (\ref{t}) and also (\ref{t1}) and the results have been compared. The results are different as figure 6 presents. One can adjust period of oscillation by using an effective potential parameter $\epsilon_{effective}$. But general behaviour of oscillation in two models are not adjustable at all. It is clear because presented model comes from different model of adding the potential to the system. But comparing  of two models show that the results are acceptable with a very good approximation.
\begin{figure}[htbp]
  \begin{center}
    \leavevmode
 \epsfxsize=11cm   \epsfbox{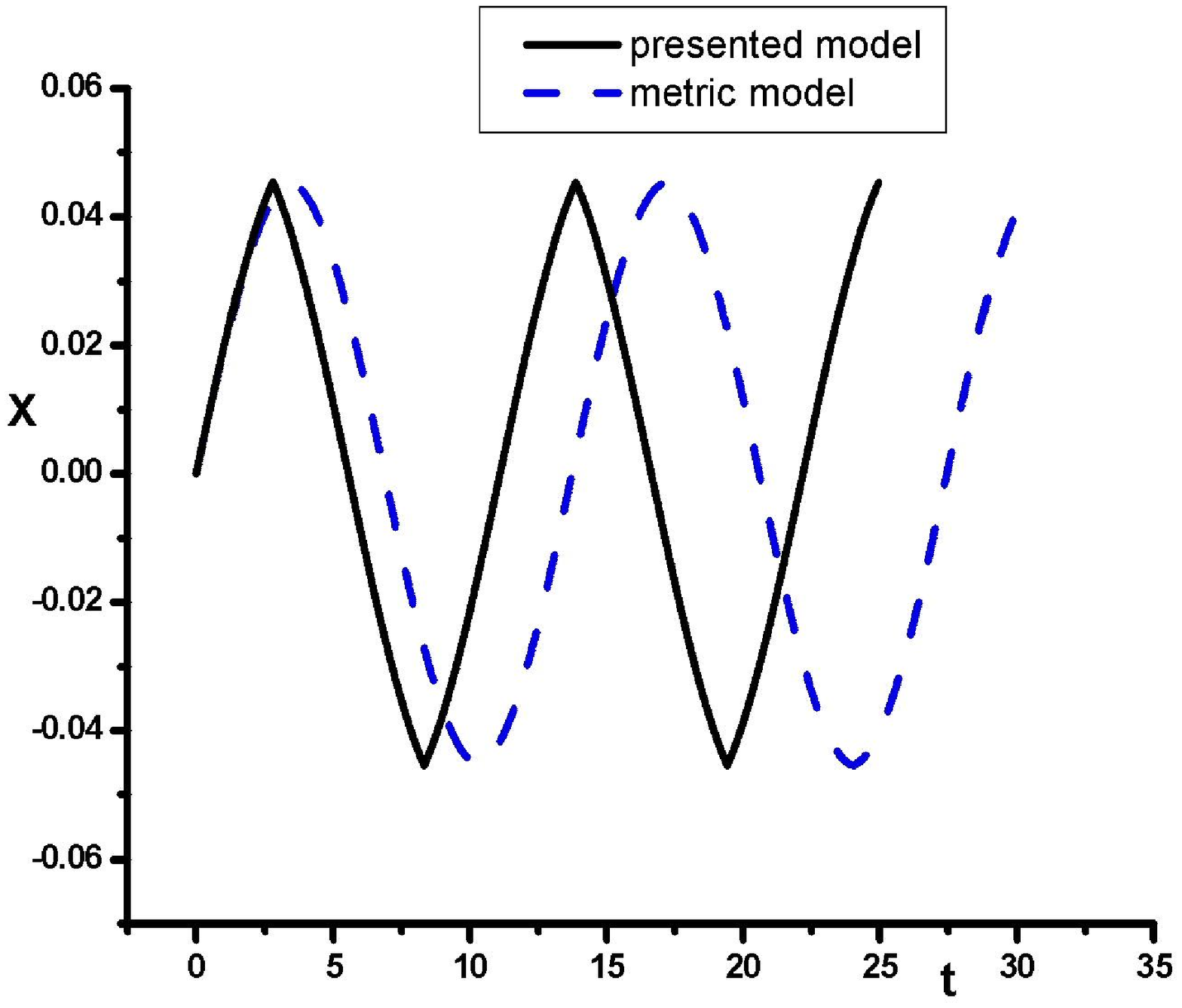}
  \end{center}
  \caption{Soliton trajectory as a function of time for a soliton with initial velocity $\dot{X}_{0}=0.02$ and initial position $X=0$ during the interaction with potential well $\epsilon=-0.3$ simulated with presented model (solid line) and metric model of Ref. \cite {r7} (dash line). }
  \label{fig:fig6}
\end{figure}

\section{Conclusion and Remarks}

An analytical model for scattering of sine-Gordon solitons from delta function potential barriers and also potential wells has been presented. Several features of soliton-potential characters were calculated using this model. A critical velocity for the soliton during the interaction with a potential barrier as a function of its initial conditions and the potential characters has been found. The model predicts specific relations between some functions of initial conditions and other functions of final state of the soliton after the interaction . An escape velocity has been derived for the soliton-well system. Oscillation period of a soliton in a potential well also has been investigated using this model. 

Calculated characters have been compared with the results of another analytical model. All of the simulations for soliton-barrier and soliton-well systems show the validity of presented analytic model. Therefore we can conclude that the presented collective coordinate method is able to explain most of the features of sin-Gordon soliton behaviour during the interaction with a potential. But this model (like analytical model \cite {r7}) is not able to explain fine structure of the islands of trapping in soliton-well system. This phenomenon is a very interesting features of soliton-potential systems. It is expected to find an acceptable explanation for this behaviour using a better model with suitable collective coordinate method. On the other hand using this method for investigation of other nonlinear models in potentials is an interesting subject.


\end{document}